\documentclass[12pt,a4paper]{article}
\usepackage{epsfig}
\begin{document}

\begin{flushright}
\it
Talk on ACAT2000 Workshop\\
Fermilab, October 16-20, 2000
\end{flushright}

\vspace{1cm}
\begin{center}
\large \bf
CompHEP-PYTHIA interface:\\
 integrated package for the collision events
generation based on exact matrix elements
\end{center}

\vspace{0.5cm}
\begin{center}
\large A.S.~Belyaev, E.E.~Boos, A.N.~Vologdin, M.N.~Dubinin, \\
        V.A.~Ilyin, A.P.~Kryukov, A.E.~Pukhov, A.N.~Skachkova, \\
        V.I.~Savrin, A.V.~Sherstnev, S.A.~Shichanin
\end{center}
\begin{center}
 \it Skobeltsyn Institute of Nuclear Physics, 
            Moscow State University \\
             Moscow 119899, Russia
\end{center}

\vspace{1cm}

\begin{abstract} 

CompHEP, as a partonic event generator, and PYTHIA, as a generator of final
states of detectable objects, are interfaced. Thus, integrated tool is
proposed for simulation of (almost) arbitrary collision processes at the
level of detectable particles. Exact (multiparticle) matrix elements,
convolution with structure functions, decays, partons hadronization and
(optionally) parton shower evolution are basic stages of calculations. The
PEVLIB library of event generators for LHC processes is described.

\end{abstract}

\vspace{1cm}



In the widely used generators PYTHIA \cite{pythia}, ISAJET \cite{isajet}
and HERWIG \cite{herwig} data bases of matrix elements of hard subprocesses
are built in. It means that matrix elements are stored as formulas.
Furthermore, the matrix element squared $|M|^2$  is represented by means of
some function modelling the behaviour of the integrand to get effective
Monte-Carlo integration and events generation. Thus, as one can see, mainly
$2\to 2$ subprocesses are included in these data bases.

However, the generation of events with 3, 4 and more bodies in the final
states of hard subprocesses is needed for the Tevatron, LHC and future
linear collider physics. One can note, in particular, that for such states
there is no possibility to construct simple analytical formulae to match
singular behaviour of $|M|^2$. Multidimensional phase space (4 dimensions
in the 3-body case plus 2 dimensions in case of hadron collisons for
convilution with parton distributions, 7+2 dimensions in the 4-body case
etc) with untrivial regions corresponding to the singularities of the
martrix element leads to complicated symbolic structures. Thus, one needs a
new approach to the generation of events at the partonic level.

Partonic level final states with top quarks, Higgs bosons and intermediate
vector bosons, like $Wtj$, $ttH$, $Wbb$, $ttbb$ and $tttt$,  give practical
examples of multidimensional phase space integration. Then, multiple
production of light quarks, gluons, leptons and photons also assumes the
evaluation of multiparticle final states if one is interested in effects at
high $p_T$, large invariant masses etc. This problem stands as an important
one especially if one should evaluate precisely background processes.

The singular behaviour of the phase space integrand is connected with
singular behaviour of propagators in Feynman diagrams: some masses are
extremely small (light quarks and leptons) or even zero (gluons and
photons), while other masses are of order 100 GeV ($M_W$, $M_Z$, $M_{top}$
$\ldots$) and collison energies are of hundreds or even thousands of GeV.
As a result, huge energy scale difference for the parameters involved
produces serious computational problems. One has to regularize the
integration measure to smooth the singularities (see, e.g., the discussion
in \cite{sings}).

To get more partons in final state one can exploit, for example, the
QCD {\it parton-shower} generation. However, this method is
good only for soft regimes (small $p_T$, small angles etc.) and fails in
case of hard production of these extra partons.

\vspace{0.5cm}

We propose to use programs created for automatic computation of matrix
elements, like CompHEP \cite{comphep}, GRACE \cite{grace} and MADGRAPH
\cite{madgraph}\footnote{see also programs presented on this Workshop by
T.~Ohl and C.~Papadopoulos \cite{Ohl,HELAC}.} as a tool for generation of
data base of hard subprocesses for generators like PYTHIA, ISAJET and
HERWIG.  In particular, at the step of the evaluation of hard subprocess
the phase space grid is adapted to match a singular behaviour of $|M|^2$.
So, we develop the {\it two stage} approach:

\begin{enumerate}
\item CompHEP produces cross sections and proper phase space
                 grid. This information is stored in special data
                 base, {\bf PEVLIB}, on the hard disk;
\item events generated by CompHEP are used as an input to generators
PYTHIA, ISAJET and HERWIG, for further decaying and hadronization of final
partonic states.
\end{enumerate}

In this paper we present the interface to provide an automatic input of
CompHEP events to PYTHIA. This interface is under the development still and
new options are assumed to be realized, in particular, automatic addition
of new events in the data base if the number of events already stored in
the existing sample is not enough, and the regeneration of events in case
of change of physical parameter set.

Some comments why we propose to separate the {\it matrix element} and
{\it decay/showers/hadroni\-zation} steps of the computations:

\begin{itemize}
\item it is easy to automate the interface;
\item more flexibility of the computation model is reached:
   \begin{itemize}
   \item it allows to develop/implement new options in these two steps
   independently (what corresponds to the standard theoretical approach: 
   at the "matrix element" 
   step quantum effects are evaluated, interference etc, while the second
   step corresponds to the probability processes);
   \item it gives a possibility to input partonic events in different
   programs for the second step (PYTHIA, ISAJET, HERWIG etc), as well as to
   create partonic
   events by different programs, e.g. CompHEP, GRACE, MADGRAPH
   etc. Of course the standardization of the partonic event files
   is necessary.
   \end{itemize}
\end{itemize}

\vspace{1cm}

In PYTHIA we use the subroutine {\sf PYUPEV} to input CompHEP events as an
external process.

CompHEP generates events (unweighted in v.41) and writes them to the file

\begin{center}
\sf events\_N.txt
\end{center}
where {\sf N} is the number of working session. This is the text file with
a header, where information about the subprocess is given, cross section 
value is written, and some information about the beams is presented (in
particular what PDF set was used and what is the QCD scale). Then, events
are written (one event - one line).

Then, a command

\begin{center}
\sf mixPEV 
\end{center}
is used to mix several subprocesses in one event flow according to their
relative weight, $\sigma_i/\Sigma_j\sigma_j$, where $\sigma_i$ is the cross
section of the $i$-th subprocess. This command, {\sf mixPEV}, randomizes
also the position of events from different subprocesses in the final event
flow.

As a result of the command {\sf mixPEV} the file {\sf Mixed.PEV} with the
events from mixed subprocesses is created. In this file headers of all
subprocesses contributed are listed in the beginning and then events are
written. In the end of each line (event) the information about color flows
is given, allowing the user to switch on, for example, Lund Fragmentation
Model or other models using color flows in the $N_c\to \infty$ limit. This
new option is realized in CompHEP v.41. The event line includes also: the
number of subprocess, to which the event belongs, and components of the
particles momenta (for in-coming particles only their z-components). 

When the command {\sf mixPEV} is completed the protocol file, {\sf
Prt.PEV}, is created. 

\vspace{1cm}

The CompHEP code is available from the following addresses:

\begin{center}
\sf http://theory.npi.msu.su/comphep/ 
\end{center}
\begin{center}
/afs/cern.ch/cms/physics/COMPHEP/V\_41.10.tar.gz
\end{center}

The CompHEP-PYTHIA interface code and command {\sf mixPEV} are available from:

\begin{center}
/afs/cern.ch/cms/physics/PEVLIB/cpyth
\end{center}
where one can find the Fortran code in directory {\sf interf45} (for
interface with PYTHIA 5.7/JETSET 7.4) and {\sf interf46} for interface with
PYTHIA v.6.x. The code for the command {\sf mixPEV} is placed in the
directory {\sf c\_source} (the {\sf mix.c} file).

One should compile Fortran routine from the directory {\sf interf46} (or
correspondingly from {\sf interf45}) and link with the PYTHIA object code.
For some platforms a user should comment the dummy routine {\sf PYUPEV} in the
original PYTHIA code.

For more details on the PYTHIA switches and other comments we refer the
reader to the file {\sf main.f}, where many comment lines are given with
corresponding explanations.

\vspace{1cm}

Using the CompHEP-PYTHIA interface we create the
PEVLIB library of partonic events for LHC processes. Each process is stored
in the subdirectory 
\begin{center}
/afs/cern.ch/cms/physics/PEVLIB/
\end{center}
In each process subdirectory there is the {\sf README} file where some details
are given concerning the partonic events computed.

Among processes already stored in PEVLIB are the following:
\begin{itemize}
\item {\sf SingleTop} (see details about the number of events generated and
subprocesses in the corresponding {\sf README} file);
\item $(H\to \tau^+\tau^-)+2tagjets$. Here electroweak background events
can be found (about $10^5$ with $Z$ off-shell and $4\cdot 10^5$ with $Z$
on-shell), and QCD background (about 65000 events);
\item $(H\to b\bar b)+t\bar t$. Here signal events are generated for
$M_H=100$, $115$, $120$ and $130$ GeV, as well as the events for background
processes $t\bar t b\bar
b$ (about 300000 events) and $t\bar t +2jets$
(about 2 millions events);
\end{itemize}

\vspace{1cm}
This work was supported by the INTAS-CERN 99-0377 grant and by the Programme
"University of Russia" (grant 990588), and done in the framework of
the CPP Collaboration \cite{CPP}.

\nocite{*}
\bibliographystyle{aipproc}

\begin{thebibliography}{99}

\bibitem{pythia} 
   T.~Sjostrand, Comput. Phys. Comm. {\bf 82} (1994) 74.

\bibitem{isajet}  
    F.E.Paige et al., {\it ISAJET 7.40: a Monte Carlo event generator for $pp$,
    $\bar p p$, and $e^+e^-$ reactions}, BNL-HET-98-39, Oct 1998,
    hep-ph/9810440.

\bibitem{herwig}  
   G.Marchesini  et al., {\it HERWIG VERSION 5.9.}, hep-ph/9607393.

\bibitem{sings} 
   V.A.~Ilyin, D.N.~Kovalenko and A.E.~Pukhov, Int. J. Mod. Phys. 
   {\bf C7} (1996) 761.

   D.~Kovalenko and A.~Pukhov,  Nucl. Instr. and Meth. {\bf A389} (1997) 299.

\bibitem{comphep}
   A.~Pukhov et al, {\it CompHEP - a package for evaluation of Feynman
   diagrams  and integration over multi-particle phase space. User's manual
   for version 33}, Preprint INP MSU 98-41/542, hep-ph/9908288.

\bibitem{grace}  
   T.~Ishikawa et al., {\it GRACE manual}, KEK Report 92-19, February 1993.

\bibitem{madgraph} 
   T.~Stelzer and W.F.~Long, Comput. Phys. Commun. {\bf 81} (1994).

\bibitem{Ohl} 
   T.~Ohl, {\it O'Mega: An Optimizing Matrix Element Generator}, hep-ph/0011243.

   T.~Ohl, {\it O'Mega\&WHIZARD: Monte Carlo Event Generator Generation For
   Future Colliders}, hep-ph/0011287.

\bibitem{HELAC}
   A.~ Kanaki and C.G.~Papadopoulos, 
   {\it HELAC-PHEGAS: automatic computation of helicity amplitudes 
   and cross sections}, hep-ph/0012004; Comput. Phys. Commun. {\bf 132} (2000)
   306.

\bibitem{CPP}
   {\sf http://wwwlapp.in2p3.fr/cpp/cpp.html}

\end{thebibliography}

\end{document}